\documentclass[apj]{emulateapj}
\usepackage{graphics}
\usepackage{alltt}
\usepackage{verbatim}

\def\spose#1{\hbox to 0pt{#1\hss}}
\def\simlt{\mathrel{\spose{\lower 3pt\hbox{$\mathchar"218$}}
    \raise 2.0pt\hbox{$\mathchar"13C$}}}
\def\simgt{\mathrel{\spose{\lower 3pt\hbox{$\mathchar"218$}}
    \raise 2.0pt\hbox{$\mathchar"13E$}}}

\newcommand{\oiii}{\mbox{[\ion{O}{3}]} $\,$}
\newcommand{\oiiiw}{\mbox{[\ion{O}{3}] $\lambda$5007} $\,$}

\newcommand{\ha}{\mbox{H$\alpha$} $\,$}

\newcommand{\oiiihb}{\mbox{[\ion{O}{3}]}/{\mbox{H$\beta$} $\,$}}

\newcommand{\niiha}{\mbox{[\ion{N}{2}]}/{\mbox{H$\alpha$} $\,$}}
\newcommand{\niihan}{\mbox{[\ion{N}{2}]}/{\mbox{H$\alpha$}}}

\slugcomment{Published in ApJ}
\shortauthors{Comerford et al.}
\shorttitle{}

\begin{document}

\title{An Excess of AGNs Triggered by Galaxy Mergers in MaNGA Galaxies of \\ Stellar Mass $\sim10^{11}$ $M_{\odot}$}

\author{Julia M. Comerford\altaffilmark{1}, Rebecca Nevin\altaffilmark{2}, James Negus\altaffilmark{1}, R. Scott Barrows\altaffilmark{1}, Michael Eracleous\altaffilmark{3}, Francisco M\"{u}ller-S\'{a}nchez\altaffilmark{4}, Namrata Roy\altaffilmark{5}, Aaron Stemo\altaffilmark{6}, Thaisa Storchi-Bergmann\altaffilmark{7}, Dominika Wylezalek\altaffilmark{8}}

\affil{$^1$Department of Astrophysical and Planetary Sciences, University of Colorado, Boulder, CO 80309, USA}

\affil{$^2$Fermi National Accelerator Laboratory, PO Box 500, Batavia, IL 60510, USA}

\affil{$^3$Department of Astronomy and Astrophysics and Center for Gravitational Wave Physics, The Pennsylvania State University, 525 Davey Lab, University Park, PA 16803, USA}

\affil{$^4$Physics Department, University of Memphis, Memphis, TN 38152, USA}

\affil{$^5$Center for Astrophysical Sciences, Department of Physics and Astronomy, Johns Hopkins University, Baltimore, MD, 21218, USA}

\affil{$^6$Department of Physics and Astronomy, Vanderbilt University, Nashville, TN 37235, USA}

\affil{$^7$Departamento de Astronomia, Universidade Federal do Rio Grande do Sul, IF, CP 15051, 91501-970 Porto Alegre, RS, Brazil}

\affil{$^8$Zentrum f\"{u}r Astronomie der Universit\"{a}t Heidelberg, Astronomisches Rechen-Institut, M\"{o}nchhofstr, 12-14 69120 Heidelberg,
Germany}

\begin{abstract}

To facilitate new studies of galaxy merger driven fueling of active galactic nuclei (AGNs), we present a catalog of 387 AGNs that we have identified in the final population of over $10,000$ $z<0.15$ galaxies observed by the SDSS-IV integral field spectroscopy survey Mapping Nearby Galaxies at Apache Point Observatory (MaNGA).  We selected the AGNs via mid-infrared {\it WISE} colors, {\it Swift}/BAT ultra hard X-ray detections, NVSS and FIRST radio observations, and broad emission lines in SDSS spectra.  By combining the MaNGA AGN catalog with a new SDSS catalog of galaxy mergers that were identified based on a suite of hydrodynamical simulations of merging galaxies, we study the link between galaxy mergers and nuclear activity for AGNs above a limiting bolometric luminosity of $10^{44.4}$ erg s$^{-1}$.  We find an excess of AGNs in mergers, relative to non-mergers, for galaxies with stellar mass $\sim10^{11}$ $M_{\odot}$, where the AGN excess is somewhat stronger in major mergers than in minor mergers.  Further, when we combine minor and major mergers and sort by merger stage, we find that the highest AGN excess occurs in post-coalescence mergers in the highest mass galaxies.  However, we find no evidence of a correlation between galaxy mergers and AGN luminosity or accretion rate.  In summary, while galaxy mergers overall do appear to trigger or enhance AGN activity more than non-mergers, they do not seem to induce higher levels of accretion or higher luminosities. We provide the MaNGA AGN Catalog and the MaNGA Galaxy Merger Catalog for the community here. \\

\end{abstract}  

\section{Introduction}

Nearly all massive galaxies are thought to host central supermassive black holes (SMBHs; \citealt{KO95.1,FE05.1}), and these SMBHs build up their mass by accreting gas as active galactic nuclei (AGNs).   The main avenues for driving gas accretion onto SMBHs are secular processes internal to galaxies (e.g., \citealt{LY79.1,CO85.1,BO02.2}) and tidal torques induced during galaxy mergers that cause gas to lose its angular momentum and infall to galaxy centers (e.g., \citealt{BA91.1,SP05.2,HO09.1}).  However, the relative roles and importances of these two mechanisms for driving SMBH growth remain unclear.  Observational studies disagree on fundamental questions such as whether AGNs are more often found in galaxy mergers (e.g., \citealt{CA01.2,VI11.1,EL19.1}) or not (e.g., \citealt{DU03.1,GA09.2,CI11.1}). 

The reason for the disagreement may be that the role of galaxy mergers in triggering AGNs also depends on factors such as galaxy merger mass ratio, galaxy merger stage, AGN luminosity, and accretion rate.  For example, some theoretical work has suggested that major mergers of galaxies preferentially trigger AGNs (e.g., \citealt{BA92.2,HO09.1,CA17.1}), though the observations have not yet converged on a conclusion (e.g., \citealt{GL15.1,VI17.1,MA19.1}).

Further, simulations predict that SMBH activity peaks in late-stage (often defined as, e.g., $<10$ kpc bulge separation) galaxy mergers \citep{DI05.1,SP05.2,CA15.1}, but the observations have been limited by only handfuls of known late-stage mergers hosting AGNs (e.g., \citealt{KO12.1,CO15.1,BA16.1}).  To further complicate the picture, other observations suggest that SMBH activity peaks in post-coalescence mergers (e.g., \citealt{EL13.1,SA14.1}).  

Meanwhile, galaxy simulations have also found that more luminous AGNs are preferentially triggered by galaxy merger-driven inflows of gas, while less luminous AGNs are preferentially fueled by the stochastic accretion of gas outside of mergers (e.g., \citealt{HO14.1,CA15.1}).  Currently, the observations are in tension about whether galaxy mergers preferentially trigger the most luminous AGN (e.g., \citealt{TR12.1,GL15.1}) or whether there is no correlation of mergers with AGN luminosity (e.g., \citealt{KO12.2,VI14.1}).  The question of whether galaxy mergers fuel higher accretion rates onto SMBHs also remains relatively untested (e.g., \citealt{MA19.1}).

A powerful tool for addressing these open questions on the roles of galaxy mergers in triggering AGNs at low redshifts is the Sloan Digital Sky Survey-IV (SDSS-IV) survey Mapping Nearby Galaxies at Apache Point Observatory (MaNGA), which enables spatially resolved studies of galaxies in unprecedented numbers (over 10,000 $z<0.15$ galaxies; \citealt{BU15.1,LA15.1}).  Here, we lay the groundwork for such analyses by first building a catalog of AGNs in MaNGA, which we identify with mid-infrared, X-ray, radio, and broad emission line observations.  This is an expansion of the catalog presented for an earlier MaNGA data release in \cite{CO20.1}. We also present a catalog of galaxy mergers in MaNGA that were identified based on a suite of hydrodynamical simulations of galaxy mergers, where the identifications include major and minor mergers as well as early-stage, late-stage, and post-coalescence mergers.  We combine our AGN catalog with the galaxy merger catalog to study the role of mergers in triggering AGNs as a function of merger mass ratio, merger stage, AGN luminosity, and SMBH accretion rate.  Future work will focus on the spatially-resolved impacts of mergers on galaxies, including spatially-resolved star formation rates and star formation histories.

This paper is organized as follows. In Section~\ref{manga}, we present the parent sample of MaNGA galaxies and their properties. In Section~\ref{agn}, we present the catalog of 387 AGNs in MaNGA, which we identified using {\it WISE} mid-infrared color cuts, {\it Swift}/BAT hard X-ray sources, NVSS/FIRST 1.4 GHz radio sources, and SDSS broad emission lines.  Section~\ref{mergers} describes how the galaxy mergers in MaNGA are identified.  In Section~\ref{results}, we present and discuss our results, which indicate an excess of AGNs in galaxy mergers, when compared to non-mergers, in the moderate to high mass galaxies in our sample. Finally, Section~\ref{conclusions} summarizes our conclusions.

We assume a Hubble constant $H_0 =70$ km s$^{-1}$ Mpc$^{-1}$, $\Omega_m=0.3$, and $\Omega_\Lambda=0.7$ throughout, and all distances are given in physical (not comoving) units.

\section{MaNGA Galaxy Sample}
\label{manga}

MaNGA is an SDSS-IV integral field spectroscopy survey of $0.01<z<0.15$ (average redshift $z=0.03$) galaxies, and its final catalog consists of over 10,000 unique galaxies \citep{BU15.1,DR15.1,LA15.1,YA16.1,BL17.1,WA17.1,AB22.1}.  MaNGA uses $2^{\prime\prime}$ fibers grouped into hexagonal bundles, which range in diameter from $12\farcs5$ to $32\farcs5$.  The observations span 3600 - 10,300 \AA \, with a spectral resolving power of $R\sim2000$.  The PSF FWHM is $2\farcs5$, corresponding to physical resolutions of 0.5 kpc to 6.5 kpc for its redshift range.  MaNGA targets galaxies with stellar masses $>10^9$ $M_\odot$, and the mass distribution is designed to be flat \citep{YA16.1}.  In addition, the MaNGA survey was designed to spectroscopically map galaxies out to at least 1.5 times the effective radius, and the typical MaNGA galaxy is mapped out to a radius of $\sim15$ kpc.  

Here, we use the final catalog of galaxies in MaNGA, which was released in the eleventh MaNGA Product Launch (MPL-11; \citealt{AB22.1}).  Our analyses rely on galaxy properties provided in the Pipe3D Value Added Catalog \citep{SA16.1,SA18.2}, including redshift and galaxy stellar mass. 

\section{MaNGA AGN Catalog}
\label{agn}

We provide here a catalog of 387 AGN identifications in the full, final population of MaNGA galaxies.  The properties presented in our catalog are shown in Table~\ref{tbl-agn}.  We select the AGNs in a similar fashion to \cite{CO20.1}, which searched for AGNs in the 6261 MaNGA galaxies in MPL-8.

We select AGNs using four different data sets, where each data set has full coverage of the MaNGA field and therefore could potentially detect AGNs in all MaNGA galaxies.   The four approaches that we use are {\it WISE} mid-infrared color cuts, {\it Swift}/BAT hard X-ray sources, NVSS/FIRST 1.4 GHz radio sources, and SDSS broad emission lines. We outline each approach below.

\begin{deluxetable*}{lll} 
\tablewidth{0pt}
\tablecolumns{3}
\tablecaption{Data Fields in the MaNGA AGN Catalog} 
\tablehead{
\colhead{No.} &
\colhead{Field} & 
\colhead{Description} 
}
\startdata 
1 & MANGA\_ID & Galaxy identifier assigned by MaNGA \\
2 & RA & Right ascension of MaNGA galaxy [J2000, decimal degrees] \\
3 & DEC & Declination of MaNGA galaxy [J2000, decimal degrees] \\
4 & Z & Spectroscopic redshift of galaxy \\
5 & WISE\_AGN & Whether AGN was selected in {\it WISE} [Boolean] \\
6 & LOG\_LBOL\_WISE & Log bolometric luminosity of {\it WISE} AGN [erg s$^{-1}$] \\
7 & LOG\_LBOL\_WISE\_ERR & Log bolometric luminosity error of {\it WISE} AGN [erg s$^{-1}$] \\
8 & BAT\_AGN & Whether AGN was selected in BAT [Boolean] \\
9 & LOG\_LBOL\_BAT & Log bolometric luminosity of BAT AGN [erg s$^{-1}$] \\
10 & LOG\_LBOL\_BAT\_ERR & Log bolometric luminosity error of BAT AGN [erg s$^{-1}$] \\
11 & RADIO\_AGN & Whether AGN was selected in NVSS/FIRST [Boolean] \\
12 & RADIO\_CLASS & Quasar-mode (HERG) or radio-mode (LERG) \\
13 & LOG\_LBOL\_RADIO & Log bolometric luminosity of NVSS/FIRST AGN [erg s$^{-1}$] \\
14 & LOG\_LBOL\_RADIO\_ERR & Log bolometric luminosity error of NVSS/FIRST AGN [erg s$^{-1}$] \\
15 & BROAD\_AGN & Whether AGN was selected by broad lines [Boolean]  \\
16 & LOG\_LBOL\_BROAD & Log bolometric luminosity of broad-line AGN [erg s$^{-1}$] \\
17 & LOG\_LBOL\_BROAD\_ERR & Log bolometric luminosity error of broad-line AGN [erg s$^{-1}$] 
\enddata
\tablecomments{The MaNGA AGN Catalog is available in its entirety in fits format from the original publisher.}
\label{tbl-agn}
\end{deluxetable*}

\subsection{WISE Mid-infrared Colors} 
\label{wise}

Mid-infrared emission from the hot dust in the obscuring structure around AGNs is a useful selector for both obscured and unobscured AGNs.  Here, we select AGNs using {\it Wide-field Infrared Survey Explorer} ({\it WISE}) mid-infrared observations of the MaNGA catalog of galaxies.  {\it WISE} observed the full sky in four bands at 3.4 $\mu$m, 4.6 $\mu$m, 12 $\mu$m, and 22 $\mu$m ($W1$, $W2$, $W3$, and $W4$, respectively).  We crossmatch the AllWISE Source Catalog \citep{WR10.1,MA14.1} to the MaNGA galaxies using a matching radius of $6\farcs25$, which is the smallest radius of a MaNGA integral field unit.  As in \cite{CO20.1}, we select AGNs using the 75\% reliability criteria of $W1-W2>0.486 \, \exp \{ 0.092(W2-13.07)^2 \}$ and $W2>13.07$, or $W1-W2 > 0.486$ and $W2 \leq 13.07$ \citep{AS18.1}, which yields 130 AGNs.   

To determine the bolometric luminosity of each {\it WISE} AGN, we estimate the rest-frame 6 $\mu$m luminosity and convert it to the restframe 2-10 keV luminosity using the correlation from \cite{ST15.1}.  Then, we convert the restframe 2-10 keV luminosity to the bolometric luminosity by multiplying by a factor of 20, which is a typical bolometric correction for AGNs (e.g., \citealt{EL94.1,MA04.4}).  

\subsection{Swift/BAT Ultra Hard X-rays} 

Hard X-rays are excellent tracers of the hot X-rays that originate close to AGN accretion disks.  The {\it Swift} observatory's Burst Alert Telescope (BAT) is carrying out a uniform all-sky survey in the ultra hard X-ray (14 - 195 keV), and the 105-month BAT catalog, which includes AGN identifications \citep{OH18.1}, has already been crossmatched to Data Release 12 of the Sloan Digital Sky Survey (SDSS DR12; \citealt{AL15.2}).  SDSS DR12 includes all of the MaNGA galaxies, and we find that there are 30 BAT-identified AGNs in MaNGA.  

For these 30 BAT AGNs, we convert the published 14 - 195 keV luminosities to bolometric luminosities using the bolometric correction $L_{bol}/L_{14-195 \, \mathrm{keV}}=8.47$ \citep{RI17.2,IC19.1}.

\subsection{NVSS/FIRST 1.4 GHz Radio Sources}

Radio jet emission is another useful indicator of AGNs.  Radio AGNs in SDSS DR7 have been identified from the NRAO Very Large Array Sky Survey (NVSS; \citealt{CO98.3}) and the Faint Images of the Radio Sky at Twenty centimeters (FIRST; \citealt{BE95.1}) observations.  NVSS is a 1.4 GHz continuum survey that fully covers the sky north of a declination of -$40$ deg, while FIRST is a 1.4 GHz survey of 10,000 square degrees of the North and South Galactic Caps. Using these two surveys, \cite{BE12.1} created an SDSS DR7 radio source catalog with a flux density limit of 5 mJy, which extends down to a 1.4 GHz radio luminosity of $\sim 10^{23}$ W Hz$^{-1}$ at $z=0.1$.  SDSS DR17 includes all of the MaNGA galaxies, and we crossmatch to the \cite{BE12.1} catalog here.

In their analysis, \cite{BE12.1} separated the radio sources into the two classes of high-excitation radio galaxies (HERGs) and low-excitation radio galaxies (LERGs).  Building on a library of optical diagnostics (e.g., \citealt{CI10.3,KE06.1}), they used optical emission lines detected in the SDSS spectra to make these classifications. Then, \cite{BE12.1} further separated the radio sources as either star-forming galaxies or radio-loud AGNs.  They did so using the relationship between the 4000\AA \, break strength and the ratio of the radio luminosity to the galaxy stellar mass \citep{BE05.3}, Baldwin-Philips-Terlevich (BPT; \citealt{BA81.1}) line flux ratio diagnostics, and the relationship between \ha luminosity and radio luminosity.  By cross-matching their AGN classifications to MaNGA, we find 221 radio AGNs detected in MaNGA.

Then, we convert the 1.4 GHz integrated fluxes of the sources to 2-10 keV luminosities via the scaling relation in \cite{PA15.1}, and apply the same bolometric correction as in Section~\ref{wise}.

\subsection{Broad Emission Lines} 

Broad Balmer emission lines trace Type 1 AGNs, as the broad lines are produced in the high density, and high velocity, gas very close to the supermassive black hole (e.g., \citealt{OS91.1}).  \cite{OH15.1} analyzed the spectra of SDSS DR7 galaxies at $z<0.2$ for evidence of broad H$\alpha$ emission lines, which they used to build a catalog of Type 1 AGNs in SDSS DR7.  We crossmatch their catalog with MaNGA, since all MaNGA galaxies are included in SDSS DR7, and we find 78 broad-line AGNs in MaNGA.

We then convert the \oiiiw luminosities given in \cite{OH15.1} to bolometric luminosities via the scaling relation of \cite{PE17.1}.

\subsection{Comparison of AGN Selectors}
\label{agn_comparison}

\begin{figure}[!t] 
\centering
\includegraphics[width=8.5cm]{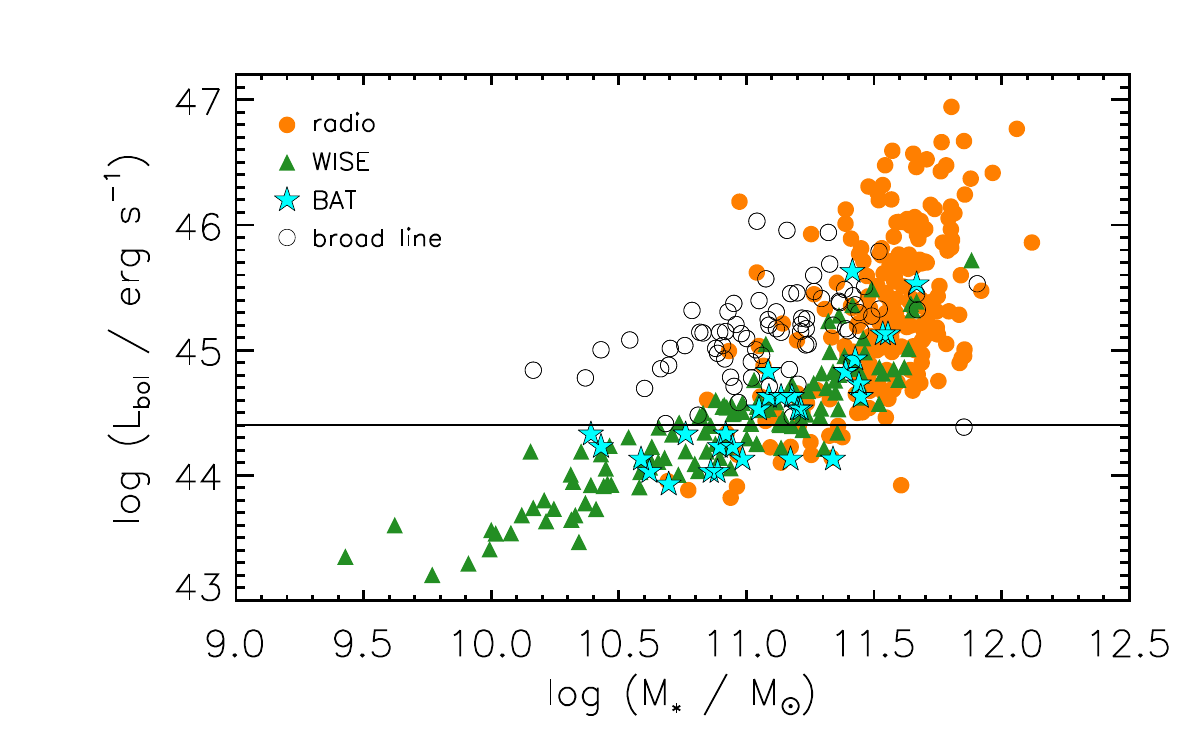}
\caption{Comparison between AGN bolometric luminosities and host galaxy stellar masses for the NVSS/FIRST radio-selected (orange filled circles), {\it WISE}-selected (green triangles), {\it Swift}/BAT-selected (cyan stars), and SDSS broad line-selected (open black circles) AGNs in MaNGA. Since each survey has a different sensitivity, we employ a limiting AGN bolometric luminosity of $\log(L_{bol}/ \mathrm{erg} \, \mathrm{s}^{-1}) > 44.4$ (horizontal line) for the analyses in this paper.  All four approaches can detect AGNs above this detection limit.}
\label{fig:agn_compare}
\end{figure}

We compare the AGN bolometric luminosities and host galaxy stellar masses of the MaNGA AGNs in Figure~\ref{fig:agn_compare}.  Several well-known trends and selection biases are apparent in our sample of AGNs.  First, the overall trend of increasing AGN bolometric luminosity with host galaxy stellar mass is due to the SMBH mass rising with the bulge stellar mass (e.g., \citealt{MA98.2}), since AGN luminosity correlates with Eddington luminosity, which correlates with SMBH mass. 

Further, there are biases in the AGN selectors.  For example, {\it WISE}-selected AGNs preferentially reside in lower stellar mass galaxies, since the infrared light from the AGN typically dominates over that from star formation in low mass galaxies, which have relatively low star formation rates compared to high mass galaxies (e.g., \citealt{ME13.2,BA21.1}).  Meanwhile, radio-selected AGNs are biased towards higher stellar mass galaxies (e.g., \citealt{BA92.3,VE01.2,KA08.1}), which could be because radio AGNs are a last phase in the evolution of AGNs when the host galaxies have already slowed their star formation (e.g., \citealt{CO20.1}).

To account for such biases, it is good practice use mass-matched control samples or to bin the AGNs in this catalog by galaxy stellar mass, as we do in the analyses that follow.

Finally, each telescope and survey that was used to detect the AGNs has a different sensitivity, which introduces a bias when combining the AGNs identified by different approaches.  To combat this bias, we employ a limiting AGN bolometric luminosity of $\log(L_{bol}/ \mathrm{erg} \, \mathrm{s}^{-1}) > 44.4$, above which all four approaches can detect an AGN.  This luminosity cutoff decreases our AGN sample size from 387 to 301.  This excludes 50\%, 48\%, 8\%, and 1\% of the {\it Swift/BAT}, {\it WISE}, radio, and broad-line AGNs, respectively. The remaining 301 AGNs, all of which are detected at $\log(L_{bol}/ \mathrm{erg} \, \mathrm{s}^{-1}) > 44.4$, are the population we use in the analyses that follow.

\begin{figure}[!t] 
\centering
\includegraphics[width=8.5cm]{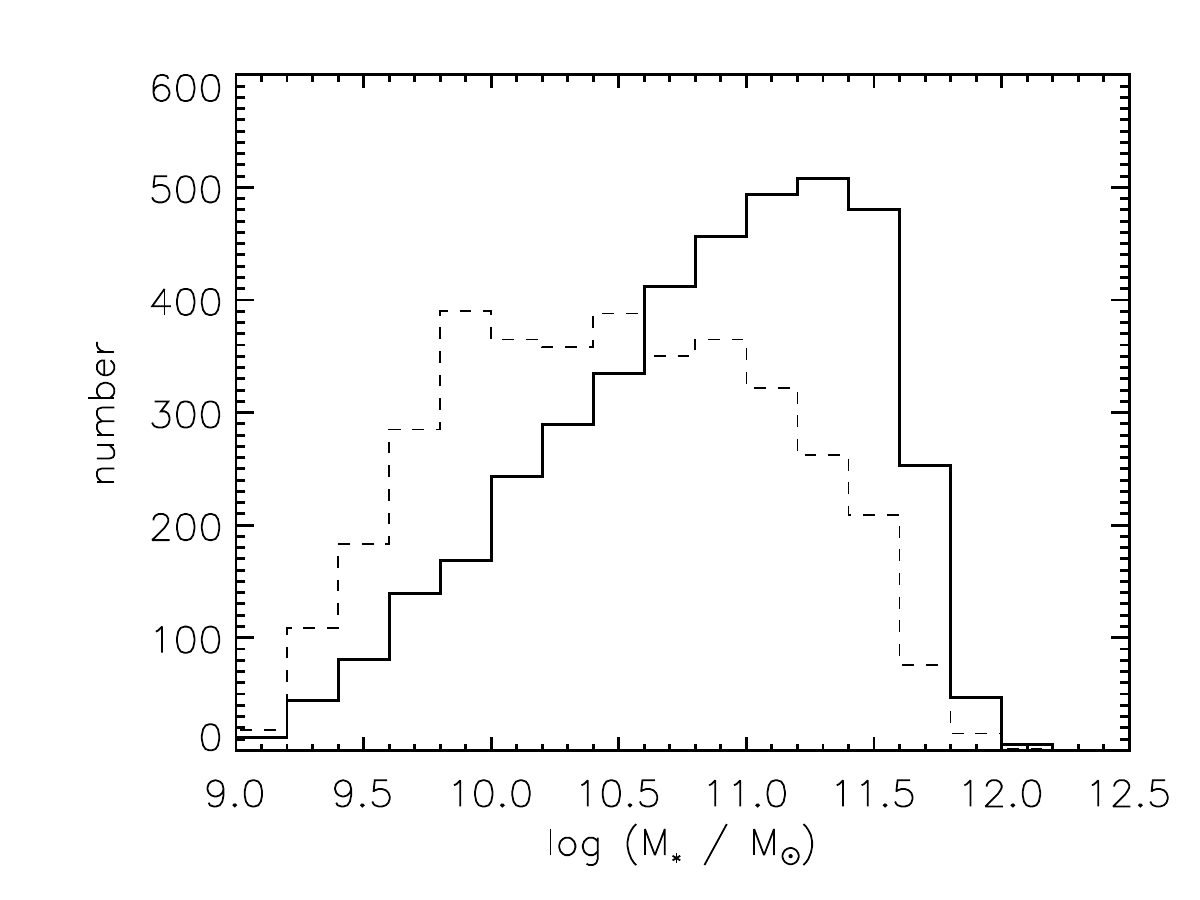}
\caption{Comparison of host galaxy stellar masses for the mergers (solid histogram) and non-mergers (dashed histogram) in MaNGA.} 
\label{fig:m_compare}
\end{figure}

\section{MaNGA Galaxy Merger Catalog}
\label{mergers}

To identify the galaxy mergers in the MaNGA catalog of galaxies, we used the SDSS galaxy classifications of \cite{NE23.1}.  They identified galaxy mergers using a linear discriminant analysis of seven individual imaging predictors (Gini, M$_{20}$, concentration, asymmetry, clumpiness, S\'ersic index, and shape asymmetry) that they measured for mock SDSS $r-$band images \citep{NE19.1}.  The mock SDSS $r-$band images were created from a suite of hydrodynamical simulations of merging and isolated galaxies that were then processed with a dust radiative transfer code. \cite{NE23.1} then applied the linear discriminant analysis classifier to the $r-$band images of the SDSS DR16 photometric sample of galaxies to determine the probability that each galaxy is an early-stage major merger, early-stage minor merger,  late-stage major merger, late-stage minor merger, post-coalescence major merger, and post-coalescence minor merger. Each probability is calculated independently, and more detail is provided in \cite{NE23.1}.

Here, major mergers are defined as galaxies with mass ratios $\mu \geq 1/4$, whereas minor mergers are defined as galaxies with mass ratios $\mu < 1/4$.  Early-stage mergers are defined as mergers roughly between first pericentric passage and apocentric passage (where the average separation of the two stellar bulges drops down to 10 kpc; \citealt{NE19.1}), late-stage mergers are defined as mergers roughly between apocentric passage and coalescence, and post-coalescence mergers are defined as the stage from coalescence until 0.5 Gyr or 1 Gyr after coalescence. In this paper we use the 1 Gyr after coalescence classification for post-coalescence mergers, because, as \cite{NE23.1} point out, IllustrisTNG galaxies exhibit disturbed morphologies for up to 2.5 Gyr following a merger \citep{BI21.1}.

The accuracies of the early-stage, late-stage, and post-coalescence major merger classifications are 0.86, 0.94, and 0.90, respectively, and the precisions are 0.95, 0.97, and 0.94, respectively \citep{NE23.1}.  The merger classification methodology performs best when applied in bulk to large populations like the MaNGA sample here, rather than to individual galaxies.

First, we match the full catalog of MaNGA galaxies to the \cite{NE23.1} SDSS DR16 galaxy classification catalog using a crossmatch radius of $20^{\prime\prime}$.  We choose this radius because it corresponds to 12 kpc at the median redshift of MaNGA and 50 kpc at the highest redshift of MaNGA, and galaxies matched within 50 kpc have been well-established to be likely merging pairs (e.g., \citealt{PA02.1,EL11.1,SN17.1}).  We find that 7669 of the MaNGA galaxies (76\%) match within this radius, are photometrically clean (no photometric flags, where the three possible photometric flags are `low S/N', `outlier predictor', and `segmap'; see Section 2.2 of \citealt{NE23.1}), and have morphology classifications that we can use in our analyses. We provide the MaNGA Galaxy Merger Catalog here, which gives classification probabilities for these 7669 MaNGA galaxies.  We ultimately classify some of these galaxies as mergers and some as non-mergers, as detailed below. The data fields in the catalog are described in Table~\ref{tbl-merger}.

In our analyses, we follow the practice recommended in \cite{NE23.1} to identify mergers and non-mergers. We use the merger probability values from the marginalized analysis that corresponds to the 50th percentile of the probability distribution for each individual galaxy.  We use a probability threshold of 0.5 to distinguish between mergers and non-mergers.  The galaxy mergers are those where {\tt p\_merg\_stat\_50\_major\_merger} or {\tt p\_merg\_stat\_50\_minor\_merger} is greater than 0.5, and the corresponding non-mergers are those where {\tt p\_merg\_stat\_50\_major\_merger} and {\tt p\_merg\_stat\_50\_minor\_merger} are less than 0.5.
 
 Major mergers are those where {\tt p\_merg\_stat\_50\_major\_merger} is greater than 0.5, and {\tt p\_merg\_stat\_50\_major\_merger} is greater than {\tt p\_merg\_stat\_50\_minor\_merger}.  Similarly, minor mergers are those where {\tt p\_merg\_stat\_50\_minor\_merger} is greater than 0.5, and {\tt p\_merg\_stat\_50\_minor\_merger} is greater than {\tt p\_merg\_stat\_50\_major\_merger}.
 
 For each galaxy, there are six merger stage probabilities in total.  To determine merger stage, we compare the merger probabilities (calculated separately for major and minor mergers) for each stage (early-stage, late-stage, and post-coalescence).  For example, early-stage mergers are those where {\tt p\_merg\_stat\_50\_major\_merger\_early} or {\tt p\_merg\_stat\_50\_minor\_merger\_early} is greater than 0.5, and either {\tt p\_merg\_stat\_50\_major\_merger\_early} or {\tt p\_merg\_stat\_50\_minor\_merger\_early}  is greater than the probabilities of all the other merger stages ({\tt p\_merg\_stat\_50\_major\_merger\_late}, {\tt p\_merg\_stat\_50\_minor\_merger\_late}, {\tt p\_merg\_stat\_50\_major\_merger\_postc\_include\_coal\_1\_0}, and {\tt p\_merg\_stat\_50\_minor\_merger\_postc\_include\_coal\_1\_0}).  If none of the six merger stage probabilities exceed 0.5, then we classify the galaxy as a non-merger.  Overall, most probability values are either at the low (0) or the high (1) end of the distribution \citep{NE23.1}.
 
 We note that the merger stage classifications have more false negatives (are more contaminated by other merger stages) than the major versus minor merger classifications \citep{NE23.1}, which leads to fewer galaxies in each classification bin.  Due to small number statistics, this results in larger error bars in the merger stage analyses that follow (e.g., Section~\ref{late_stage}).
 
To account for the bias that galaxy mergers are preferentially found in higher mass galaxies (Figure~\ref{fig:m_compare}), we bin by galaxy stellar mass in the analyses that follow.

\begin{deluxetable*}{lll} 
\tablewidth{0pt}
\tablecolumns{3}
\tablecaption{Data Fields in the MaNGA Galaxy Merger Catalog} 
\tablehead{
\colhead{No.} &
\colhead{Field} & 
\colhead{Description} 
}
\startdata 
1 & ID	& SDSS ObjID from DR16 \\
2 & RA\_MaNGA & Right ascension in the MaNGA catalog [J2000, decimal degrees]\\
3 & dec\_MaNGA & Declination in the MaNGA catalog [J2000, decimal degrees]\\
4 & RA\_SDSS & Right ascension in the SDSS catalog [J2000, decimal degrees]\\
5 & dec\_SDSS & Declination in the SDSS catalog [J2000, decimal degrees]\\
6 & low S/N & Photometric flag, value is 1 if thrown. See N23 for more details on the three \\
 & & photometric flags. The `low S/N' flag is thrown when the average S/N value \\
 & & is below 2.5.\\
7 & outlier predictor	& Same as above, the `outlier predictor' flag is thrown when one or more imaging \\
& & predictors from the LDA classification are outside the range of predictor values \\
& & from the simulated galaxies. 
\\
8 & segmap & Same as above, the `segmap' flag is thrown when the segmentation map used for \\
& & imaging analysis does not include the central pixel or for when the segmentation \\
& & map extends beyond the edge of a clipped image. 
 \\
9 & p\_merg\_stat\_16\_major\_merger & Merger probability value from the marginalized analysis that corresponds to the \\
& & 16th percentile of the probability distribution for this individual galaxy for the \\
& & major merger overall (all stages) classification\\
10 & p\_merg\_stat\_50\_major\_merger	& Same as above but for the 50th percentile, or median of the probability \\
& &  distribution for the major merger overall (all stages) classification\\
11 & p\_merg\_stat\_84\_major\_merger & Same as above but for the 84th percentile for the major merger overall (all stages) \\
& & classification \\
12 & cdf\_major\_merger & Cumulative distribution function value that corresponds to the 50th percentile \\
& & probability value given above. The full details of this calculation are in N23. \\
& & Briefly, this value is useful for comparing how an individual galaxy's p\_merg\_50 \\
& & value compares to all SDSS galaxies classified using the same classification.\\
13 & p\_merg\_stat\_16\_major\_merger\_early	& Same as above but for the classification that identifies major mergers in the early \\
& & stage\\
14 & p\_merg\_stat\_50\_major\_merger\_early	 & \\
15 & p\_merg\_stat\_84\_major\_merger\_early	& \\
16 & cdf\_major\_merger\_early	 & \\
17 & p\_merg\_stat\_16\_major\_merger\_late & Same as above but for the classification that identifies major mergers in the late \\
& & stage\\
18 & p\_merg\_stat\_50\_major\_merger\_late	& \\
19 & p\_merg\_stat\_84\_major\_merger\_late	& \\
20 & cdf\_major\_merger\_late	 & \\
21 & p\_merg\_stat\_16\_major\_merger\_prec	& Same as above but for the classification that identifies major mergers in the early \\
& & and late stages (pre-coalescence)\\
22 & p\_merg\_stat\_50\_major\_merger\_prec & \\
23 & p\_merg\_stat\_84\_major\_merger\_prec	 & \\
24 & cdf\_major\_merger\_prec	& \\
25 & p\_merg\_stat\_16\_major\_merger\_postc\_include\_coal\_0.5	& Same as above but for the classification that identifies major mergers in the \\
& & post-coalescence stage, where the merger is defined to end 0.5 Gyr after coalescence\\
26 & p\_merg\_stat\_50\_major\_merger\_postc\_include\_coal\_0.5	 & \\
27 & p\_merg\_stat\_84\_major\_merger\_postc\_include\_coal\_0.5	& \\
28 & cdf\_major\_merger\_postc\_include\_coal\_0.5	 & \\
29 & p\_merg\_stat\_16\_major\_merger\_postc\_include\_coal\_1.0	& Same as above but for the classification that identifies major mergers in the \\
& & post-coalescence stage, where the merger is defined to end 1.0 Gyr after coalescence\\
30 & p\_merg\_stat\_50\_major\_merger\_postc\_include\_coal\_1.0	 & \\
31 & p\_merg\_stat\_84\_major\_merger\_postc\_include\_coal\_1.0	 & \\
32 & cdf\_major\_merger\_postc\_include\_coal\_1.0	 & \\
33 & p\_merg\_stat\_16\_minor\_merger & Same as above but for the minor merger classifications\\
34 & p\_merg\_stat\_50\_minor\_merger	& \\
35 & p\_merg\_stat\_84\_minor\_merger & \\
36 & cdf\_minor\_merger & \\
37 & p\_merg\_stat\_16\_minor\_merger\_early & \\
38 & p\_merg\_stat\_50\_minor\_merger\_early & \\
39 & p\_merg\_stat\_84\_minor\_merger\_early	& \\
40 & cdf\_minor\_merger\_early	 & \\
41 & p\_merg\_stat\_16\_minor\_merger\_late	 & \\
42 & p\_merg\_stat\_50\_minor\_merger\_late	 & \\
43 & p\_merg\_stat\_84\_minor\_merger\_late	 & \\
44 & cdf\_minor\_merger\_late	 & \\
45 & p\_merg\_stat\_16\_minor\_merger\_prec	 & \\
46 & p\_merg\_stat\_50\_minor\_merger\_prec	 & \\
47 & p\_merg\_stat\_84\_minor\_merger\_prec	 & \\
48 & cdf\_minor\_merger\_prec & \\
49 & p\_merg\_stat\_16\_minor\_merger\_postc\_include\_coal\_0.5	& \\
50 & p\_merg\_stat\_50\_minor\_merger\_postc\_include\_coal\_0.5	& \\
51 & p\_merg\_stat\_84\_minor\_merger\_postc\_include\_coal\_0.5	& \\
52 & cdf\_minor\_merger\_postc\_include\_coal\_0.5	& \\
53 & p\_merg\_stat\_16\_minor\_merger\_postc\_include\_coal\_1.0	& \\
54 & p\_merg\_stat\_50\_minor\_merger\_postc\_include\_coal\_1.0	& \\
55 & p\_merg\_stat\_84\_minor\_merger\_postc\_include\_coal\_1.0 & \\
56 & cdf\_minor\_merger\_postc\_include\_coal\_1.0 & 
\enddata
\tablecomments{More details are given for the derivation of all columns in \cite{NE23.1}.  The MaNGA Galaxy Merger Catalog is available in its entirety in fits format from the original publisher.}
\label{tbl-merger}
\end{deluxetable*}

\section{Results and Discussion}
\label{results}

\subsection{An Excess of AGNs in Galaxy Mergers of Stellar Mass $\sim10^{11}$ $M_{\odot}$} 
\label{fraction}

Overall, we find that $4.2 \pm 0.3$\% (168/3969) of galaxy mergers host AGNs and $1.8 \pm 0.2$\% (66/3700) of non-mergers host AGNs with $\log (L_{bol} / \mathrm{erg \, s}^{-1}) > 44.4$ in MaNGA.  For comparison, another study of AGN fractions in MaNGA galaxies, which used close galaxy pairs to define the merger sample, found that 23.8\% of mergers host AGNs while 14.5\% of non-mergers host AGNs \citep{FU18.1}.  The higher AGN fractions in that study are likely explained by the difference in AGN luminosity range ($41 \simlt \log (L_{bol} / \mathrm{erg \, s}^{-1}) \simlt 45$ for \citealt{FU18.1}) as well as the difference in AGN selection;  \cite{FU18.1} selected AGN via a combination of the \oiiihb vs. \niiha BPT diagnostic \citep{BA81.1} and the WHAN (\ha equivalent width vs. \niihan; \citealt{CI10.3}) diagnostic.  Such AGN diagnostics have known issues with misclassification, since the relevant emission line ratio values can be changed by emission from shocks, young hot stars, and evolved hot stars (e.g., \citealt{RI11.1,KE13.1,AL23.1}).  Our AGN classifications via mid-infrared, hard X-ray, radio, and broad emission lines are inherently more conservative.

\begin{figure}[!t]
\centering
\includegraphics[width=8.5cm]{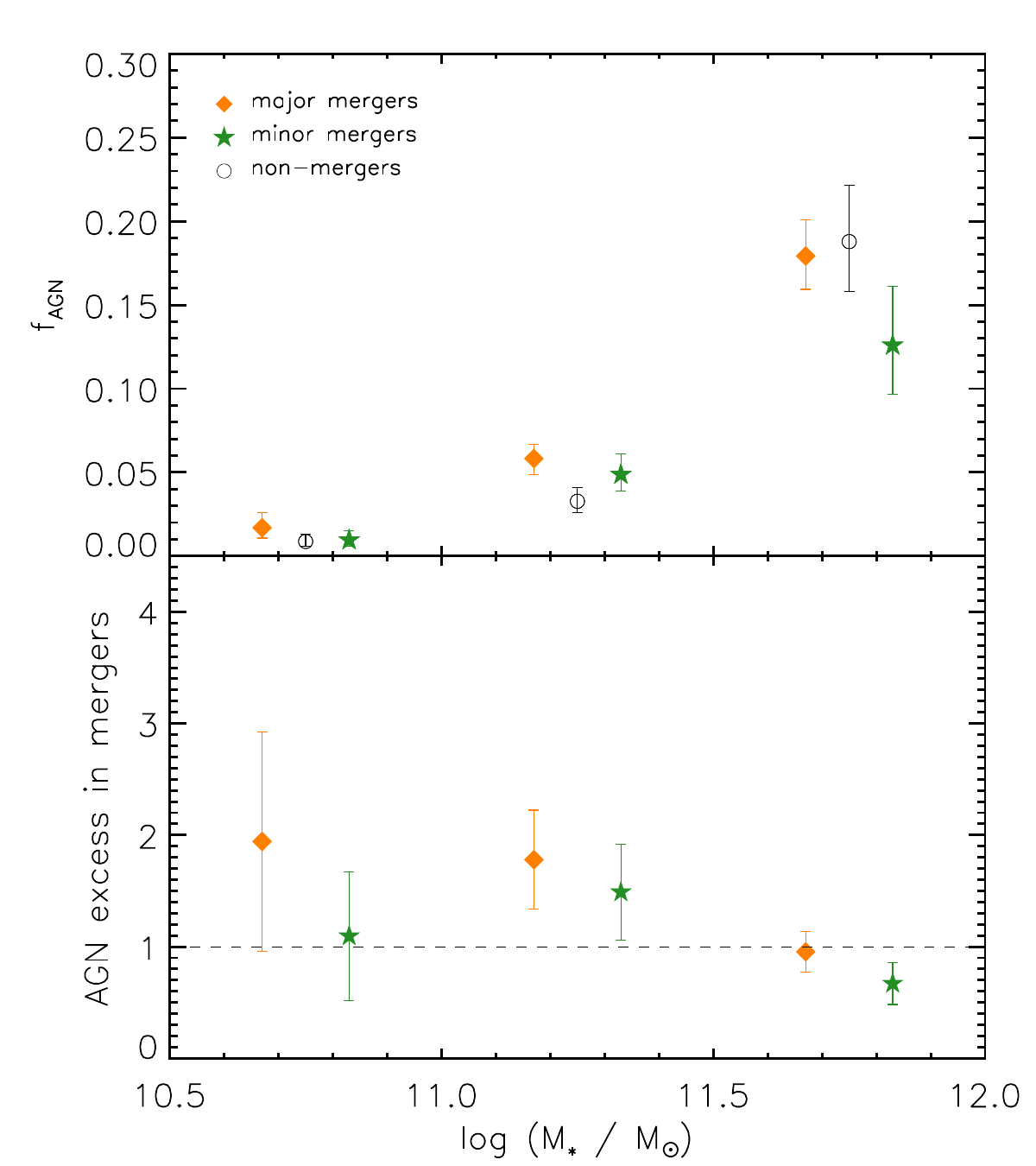}
\caption{The fraction of galaxies that host AGNs and the AGN excess as a function of galaxy stellar mass, for different mass ratio mergers.  Top: the fraction of galaxies that host AGNs increases with galaxy stellar mass for major mergers (filled orange diamonds), minor mergers (filled green stars), and non-mergers (open black circles).  The data on the x-axis are binned by 0.5 dex, and the points are offset on the x-axis for clarity.  Bottom: the AGN excess, which is defined as the fraction of mergers that host AGNs over the fraction of non-mergers that host AGNs.  The dashed line illustrates a ratio of 1, which means no difference between the fraction of mergers that host AGNs and the fraction of non-mergers that host AGNs.  We observe AGN excesses for major mergers and minor mergers in the galaxy stellar mass bin $11 < \log (M_* / M_\odot) < 11.5$.}
\label{fig:fagn}
\end{figure}

Since the AGN fraction and merger fraction are correlated with host galaxy stellar mass, we also examine the fraction of AGNs in mergers $f_{AGN}$ as compared to non-mergers within the same galaxy stellar mass bins.  Using galaxy stellar mass bins of 0.5 dex centered at $\log (M_* / M_\odot) = [10.75, 11.25, 11.75]$, we measure $f_{AGN}$ in each mass bin for the major mergers, minor mergers, and non-mergers.  We reproduce the well-known trend of the AGN fraction increasing with host galaxy stellar mass, for each galaxy type (Figure~\ref{fig:fagn}, top panel).  In each galaxy stellar mass bin, we find that the AGN fraction trends higher in the major mergers than in the minor mergers.   By comparing the fraction of AGNs in mergers to the fraction of AGNs in mass-matched non-mergers, we find that there is an AGN excess in major mergers and in minor mergers in the galaxy stellar mass bin $11 < \log (M_* / M_\odot) < 11.5$ (Figure~\ref{fig:fagn}, bottom panel).  The excess for major mergers is a factor of 1.8, and the excess for minor mergers is a factor of 1.5.

This effect was predicted in cosmological hydrodynamical simulations that examined the AGN fraction in major mergers at a range of redshifts \citep{MC20.1}.  That study only used AGNs with bolometric luminosities $L_{bol} \geq 10^{43}$ erg s$^{-1}$, which encompasses all of our MaNGA AGNs.  In their lowest redshift bin ($0 < z < 1$) they found the greatest excess of AGNs triggered by major mergers in moderate mass galaxies ($\log (M_* / M_\odot) \sim 10 - 11$), and no AGN excess at higher galaxy masses.  Our observations match these predictions. 

Our results are also broadly consistent with the AGN excesses observed in other studies at similar redshift ranges.  A MaNGA study of galaxy pairs, where the AGNs were selected by BPT and WHAN diagnostics, found that for 10 kpc separation galaxy pairs the rate of stochastic fueling was enhanced by a factor of 1.3 \citep{ST23.1}.  They found no strong correlation of AGN excess with galaxy merger mass ratio, although they calculated their mass ratios using the stellar mass within the central 1 kpc of the galaxies instead of the total stellar mass of the galaxies, which prevents a direct comparison with our results.  Further, in a study of mid-infrared selected AGNs and BPT-selected AGNs in galaxy mergers (identified by a deep learning convolutional neural network) in SDSS ($0.005 < z < 0.1$),  \cite{GA20.1} found an AGN excess by a factor of up to 1.8 in mergers. In another SDSS study at $0.01 < z < 0.20$, \cite{EL11.1} (and, similarly, \citealt{EL13.1}) used AGNs identified via BPT diagnostics and galaxy mergers identified via close pairs to uncover an AGN excess by a factor of up to 2.5 in mergers.  They also found the strongest AGN excess in equal-mass galaxy pairs.  With the close pair sample in SDSS, \cite{SA14.1} found a similar excess using mid-infrared selected AGNs.  A different mid-infrared analysis of dusty AGNs in SDSS ($0.01 < z \leq 0.08$), which used visual identification of mergers, found an AGN excess by a factor of $2-7$ in mergers \citep{WE17.1}.

\subsection{An Excess of AGNs in Post-coalescence Galaxy Mergers of Stellar Mass $\sim10^{12}$ $M_{\odot}$}
\label{late_stage}

\begin{figure}[!t]
\centering
\includegraphics[width=8.5cm]{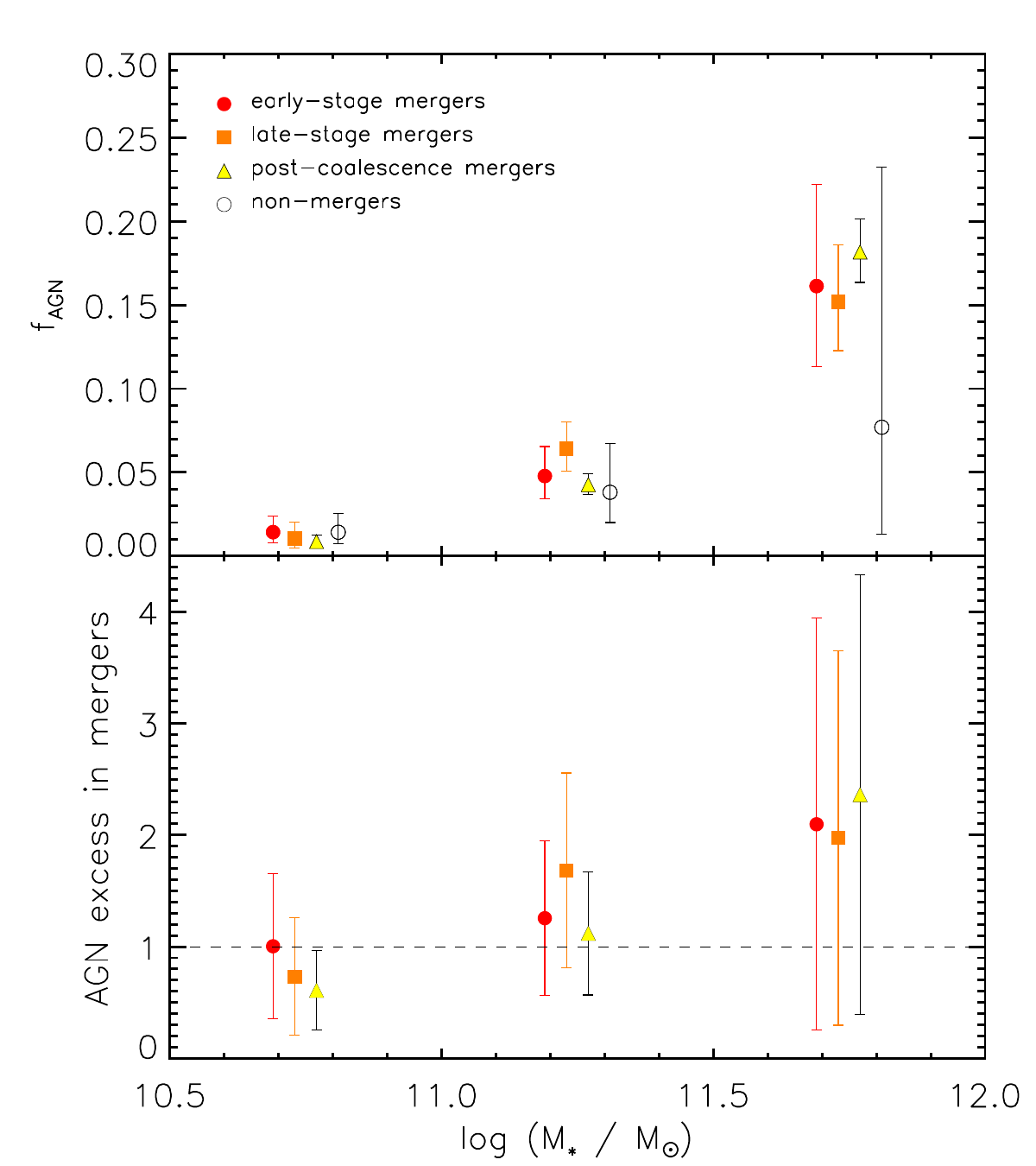}
\caption{The fraction of galaxies that host AGNs and the AGN excess as a function of galaxy stellar mass, for different merger stages.  Top: the fraction of galaxies that host AGNs increases with galaxy stellar mass for early-stage mergers (filled red circles), late-stage mergers (filled orange squares), post-coalescence mergers (filled yellow triangles), and non-mergers (open black circles).  The data on the x-axis are binned by 0.5 dex, and the points are offset on the x-axis for clarity.  Bottom: the AGN excess, which is defined as the fraction of mergers that host AGNs over the fraction of non-mergers that host AGNs.  The dashed line illustrates a ratio of 1, which means no difference between the fraction of mergers that host AGNs and the fraction of non-mergers that host AGNs.  The highest AGN excess we observe is for post-coalescence galaxy mergers in the most massive galaxy stellar mass bin ($11.5 < \log (M_* / M_\odot) < 12$), though it is still consistent with no excess to within the error bars.} 
\label{fig:fagn_mass_stage}
\end{figure}

Hydrodynamical simulations of galaxy mergers have predicted that AGN activity in mergers is preferentially triggered towards the end of the merger sequence, when the merger is funneling the most gas onto the AGNs (e.g., \citealt{VA12.1,BL13.1,CA17.1,RO19.1,MC20.1}).  Here, we test that prediction using the same galaxy stellar mass bins as in Section~\ref{fraction}, and the same limiting AGN bolometric luminosity of $\log(L_{bol}/ \mathrm{erg} \, \mathrm{s}^{-1}) > 44.4$ (Section~\ref{agn_comparison}).  As Figure~\ref{fig:fagn_mass_stage} shows, our results suggest that post-coalescence mergers have an AGN excess by a factor of 2.4, when compared to the AGN fraction in non-mergers, in the highest galaxy stellar mass bin that we consider ($11.5 < \log (M_* / M_\odot) < 12$).  This is the highest AGN excess that we find of any merger stage in any galaxy stellar mass bin, though it is still consistent, within the error bars, with no excess.

Our measured AGN excess in post-coalescence mergers is broadly consistent with results from galaxy pair observations, where the smallest separation galaxy pairs trace the latest stages of galaxy mergers.   A MaNGA analysis of AGNs in galaxy pairs also found that the fraction of AGNs increases with decreasing pair separation \citep{ST23.1}. In that study, the AGNs were selected by BPT and WHAN diagnostics. Meanwhile, in another MaNGA study that classified AGNs by BPT and WHAN diagnostics, and galaxy merger stages by a combination of kinematics and morphology, \cite{JI21.1} found no evolution of the AGN fraction across different merger sequences.  The discrepancies between these MaNGA studies may be due to differing volume corrections and aperture sizes (e.g., \citealt{ST23.1}). Studies of X-ray and mid-infrared selected AGNs in galaxy pairs also found that the fraction of AGNs increases as the pair separation decreases (e.g., \citealt{KO12.1,BA17.1,ST21.1,BA23.1}), though these studies did not search for a galaxy mass dependence.  

\subsection{Higher Luminosity AGNs Are Not Preferentially Found in Mergers, Regardless of Merger Mass Ratio or Merger Stage}

\begin{figure}[!t]
\centering
\includegraphics[width=8.5cm]{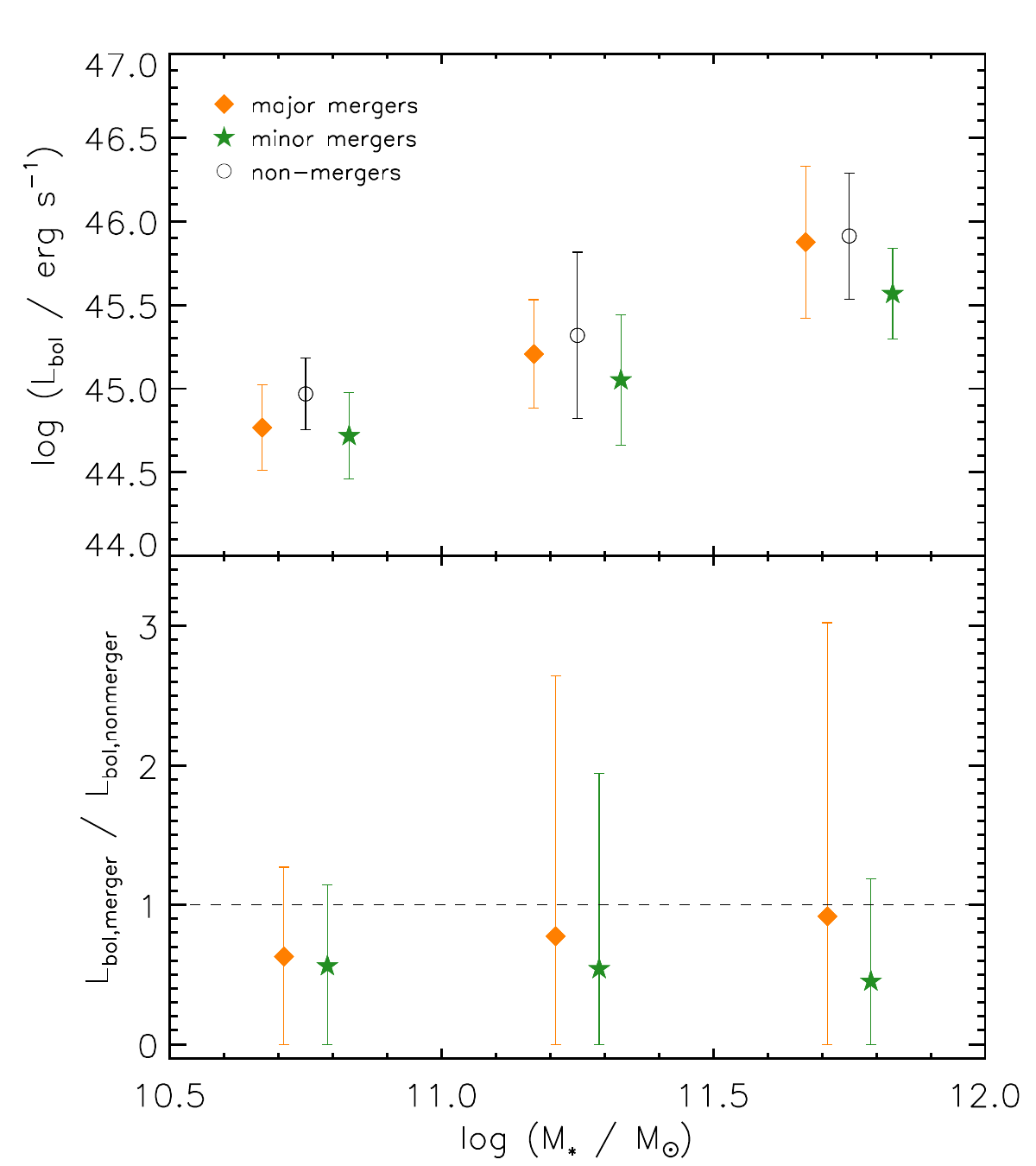}
\caption{Top: mean bolometric luminosity of AGNs in major mergers (filled orange diamonds), minor mergers (filled green stars), and the mass-matched sample of AGNs in non-mergers (open black circles).  Bottom: the ratio of the mean AGN bolometric luminosity in mergers to the mean AGN bolometric luminosity in mass-matched non-mergers, where the dashed horizontal line illustrates a ratio of 1.  The AGNs analyzed have bolometric luminosities down to a limiting luminosity of $\log(L_{bol}/ \mathrm{erg} \, \mathrm{s}^{-1}) > 44.4$.  The data on the x-axis are binned by 0.5 dex, and the points are offset on the x-axis for clarity.   While the well-known trend of increasing AGN luminosity with galaxy stellar mass is seen in all galaxy morphologies, there is no evidence for a luminosity dependence in AGN triggering in mergers.}
\label{fig:lbol_mass}
\end{figure}

There is longstanding disagreement about whether galaxy mergers, or perhaps only certain types of mergers such as major mergers or post-coalescence mergers, more effectively drive gas onto SMBHs and trigger the most luminous AGNs (see, e.g., \citealt{ST19.1} for a review).  First, we test this by comparing the AGN bolometric luminosities in major mergers and minor mergers to non-mergers within the same mass bins.  We use the same galaxy stellar mass bins as in Section~\ref{fraction} and the same limiting AGN bolometric luminosity of $\log(L_{bol}/ \mathrm{erg} \, \mathrm{s}^{-1}) > 44.4$ (Section~\ref{agn_comparison}) for our analysis.  The top panel of Figure~\ref{fig:lbol_mass} shows the known trend of AGN luminosity increasing with galaxy stellar mass, which we discussed in Section~\ref{agn_comparison}. More interestingly, that figure also shows that the mean AGN bolometric luminosity is roughly the same for major mergers, minor mergers, and non-mergers in the same galaxy mass bin.  We find no evidence for major mergers, or minor mergers, preferentially triggering higher luminosity AGNs.

Our result is in agreement with previous studies that used infrared and/or X-ray selectors of AGNs (e.g., \citealt{KO12.2,VI14.1,BA17.1,VI17.1,ST21.1}).  In contrast, other studies of AGNs in SDSS and in MaNGA found that the \oiii luminosities of AGNs are $\sim 0.13 - 0.5$ dex higher in mergers than in non-mergers \citep{EL13.1,JI21.1}.  These studies used BPT-based AGN identifications, which are prone to misclassifications as discussed in Section~\ref{fraction}; further, \oiii bolometric corrections depend on factors such as luminosity and the ionization state of the gas (e.g., \citealt{NE09.1}). These effects may contribute to the difference in our results.

\begin{figure}[!t]
\centering
\includegraphics[width=8.5cm]{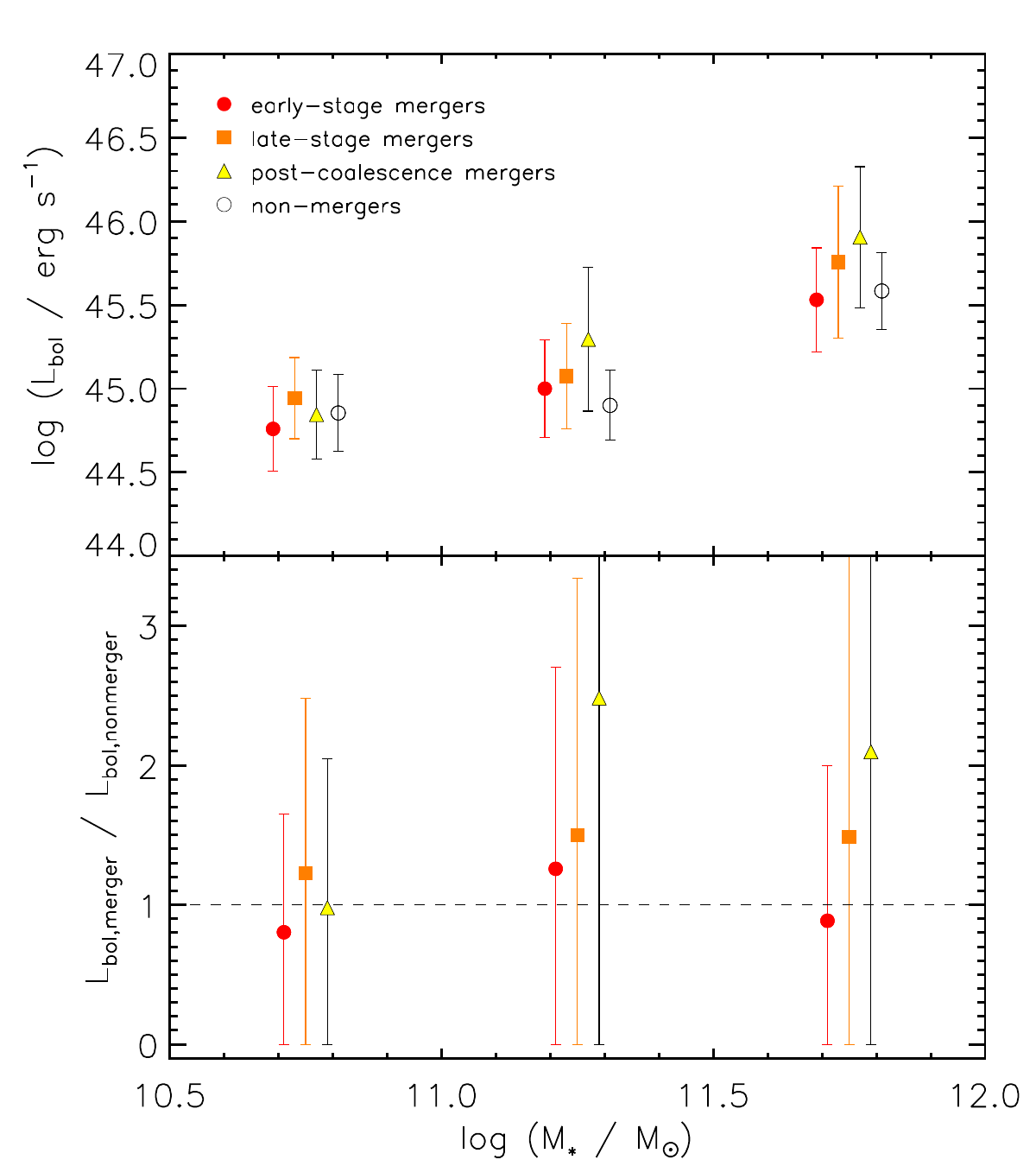}
\caption{Top: mean bolometric luminosity of AGNs in early-stage mergers (filled red circles), late-stage mergers (filled orange squares), post-coalescence mergers (filled yellow triangles), and non-mergers (open black circles).  Bottom: the ratio of the mean AGN bolometric luminosity in mergers to the mean AGN bolometric luminosity in mass-matched non-mergers, where the dashed horizontal line illustrates a ratio of 1.  The AGNs analyzed have bolometric luminosities down to a limiting luminosity of $\log(L_{bol}/ \mathrm{erg} \, \mathrm{s}^{-1}) > 44.4$.  The data on the x-axis are binned by 0.5 dex, and the points are offset on the x-axis for clarity.  While the mean AGN bolometric luminosities in post-coalescence mergers are elevated by a factor of $\sim2-3$ in the highest galaxy mass bins, they are still consistent within the errors with the mean AGN bolometric luminosities in non-mergers.}
\label{fig:lbol_mass_stage}
\end{figure}

Next, we consider the AGN luminosity as a function of merger stage.  While the AGNs in post-coalescence mergers of high stellar mass galaxies ($11 < \log (M_* / M_\odot) < 12$) seem to have bolometric luminosities that are $\sim2-3$ times larger, on average, than those of AGNs in non-mergers of the same mass range, it is not a statistically significant result (Figure~\ref{fig:lbol_mass_stage}).  If real, this hint of a correlation would match findings that post-merger galaxies have more enhanced AGN luminosities than other merger stages (e.g., \citealt{EL13.1}), but within the error bars our results are consistent with other studies that found no evolution of AGN luminosity with merger stage (e.g., \citealt{ST23.1}).

\subsection{Galaxy Mergers Do Not Trigger Higher SMBH Accretion Rates}

\begin{figure}[!t]
\centering
\includegraphics[width=8.5cm]{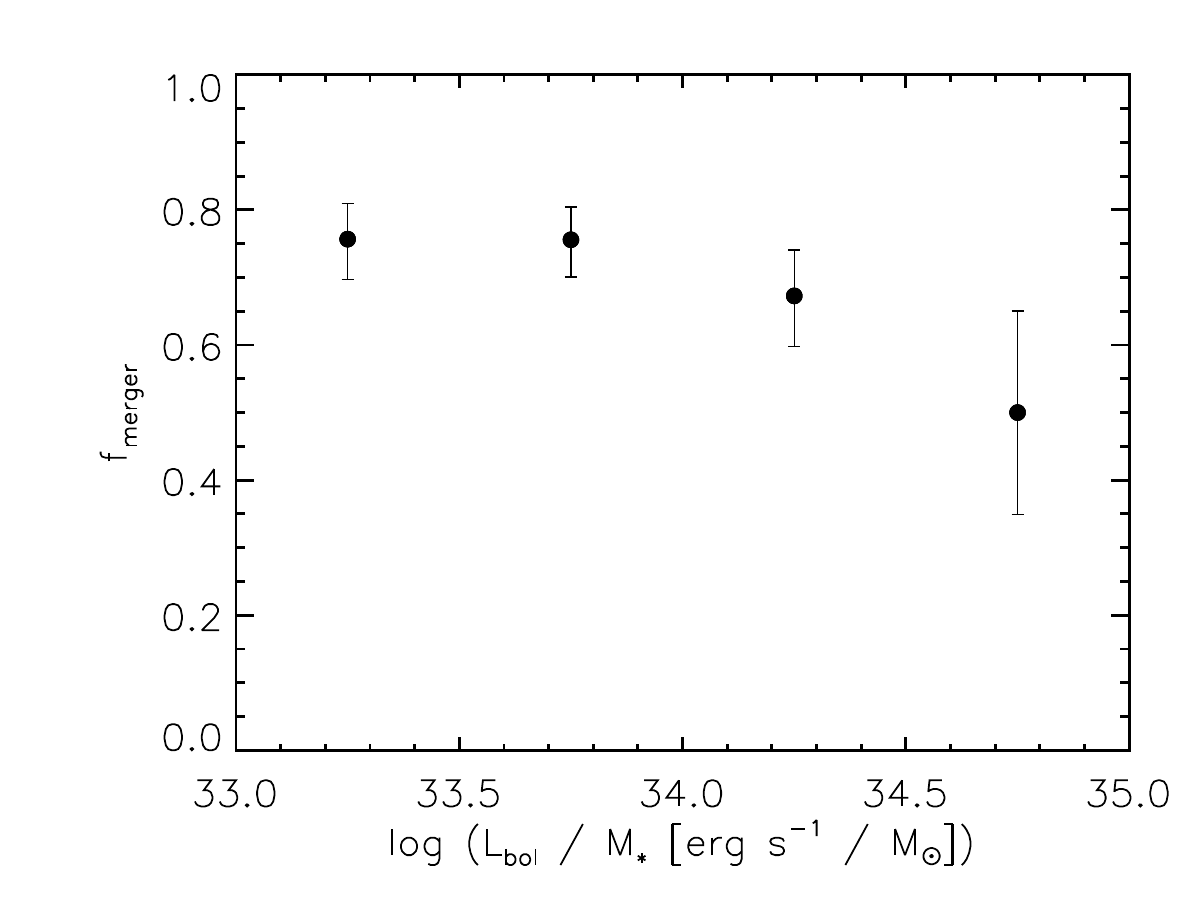}
\caption{Fraction of AGNs in mergers as a function of the ratio of AGN bolometric luminosity to the host galaxy stellar mass, which is a proxy for the Eddington ratio.  The AGNs analyzed have bolometric luminosities down to a limiting luminosity of $\log(L_{bol}/ \mathrm{erg} \, \mathrm{s}^{-1}) > 44.4$.  The data on the x-axis are binned by 0.5 dex. Within the error bars, we find no significant evidence for mergers preferentially triggering lower or higher Eddington ratios.}
\label{fig:f_lbolmass}
\end{figure}

While we find an excess of AGNs in higher-mass galaxy mergers (Section~\ref{fraction}, Section~\ref{late_stage}), we now consider whether these galaxy mergers enhance the accretion efficiencies onto the SMBHs.  Here, we use the ratio of the AGN bolometric luminosity to the host galaxy stellar mass, $L_{bol}/M_*$, as a proxy for the Eddington ratio.  Using the same limiting AGN bolometric luminosity of $\log(L_{bol}/ \mathrm{erg} \, \mathrm{s}^{-1}) > 44.4$ (Section~\ref{agn_comparison}), and bins of 0.5 dex centered at $\log [ (L_{bol}/M_*) (\mathrm{erg} \, \mathrm{s}^{-1} / M_\odot)] = [33.25, 33.75, 34.25, 34.75]$, we measure $f_{merger}$ in each bin.  The merger fraction $f_{merger}$ is defined as the number of AGNs in mergers divided by the number of AGNs, in each bin.  We find no significant trend of a dependence of $f_{merger}$ on Eddington ratio (traced by $L_{bol}/M_*$), as shown in Figure~\ref{fig:f_lbolmass}.

Our results are consistent with simulations and phenomenological models that found no significant differences in the Eddington ratios of AGNs in galaxy mergers \citep{ST16.1,WE18.1,MC20.1}.  In addition, observations of {\it WISE} AGNs in galaxy pairs also reported no significant correlation of the AGN merger fraction with $L_{bol}/M_*$  \citep{BA23.1}.  A further study of broad-line AGNs at $z\sim2$ also found that major mergers do not correlate with high accretion AGNs \citep{MA19.1}.  We conclude that galaxy mergers do not trigger higher accretion rates of material onto SMBHs for AGNs with $\log (L_{bol} / \mathrm{erg \, s}^{-1}) > 44.4$.

\section{Conclusions}
\label{conclusions}

We present identifications of 387 unique AGNs in the full, final catalog of MaNGA galaxies. We identified these AGNs by their {\it WISE} colors, {\it Swift}/BAT ultrahard X-ray sources, NVSS/FIRST radio observations, and/or broad emission lines. We combined these AGN identifications with MaNGA galaxy merger classifications, where the galaxy morphologies are derived from a linear discriminant analysis of seven individual imaging predictors (Gini, M$_{20}$ , concentration, asymmetry, clumpiness, S\'ersic index, and shape asymmetry) that were measured for mock SDSS $r-$band images of simulated merging and isolated galaxies \citep{NE23.1}.  These galaxy classifications included major and minor mergers, as well as early-stage, late-stage, and post-coalescence mergers, enabling us to examine the roles of various types and stages of mergers in triggering or enhancing AGN activity.

For the analyses in this paper, we applied a limiting AGN bolometric luminosity of $\log(L_{bol}/ \mathrm{erg} \, \mathrm{s}^{-1}) > 44.4$, above which all four approaches can detect an AGN.  This luminosity cutoff decreased our AGN sample size from 387 to 301.  Our main results for this AGN sample are summarized below.

1.  Overall, we find that $4.2 \pm 0.3$\% (168/3969) of MaNGA galaxy mergers host AGNs with $\log (L_{bol} / \mathrm{erg \, s}^{-1}) > 44.4$ and $1.8 \pm 0.2$\% (66/3700) of MaNGA non-mergers host AGNs with $\log (L_{bol} / \mathrm{erg \, s}^{-1}) > 44.4$.  

2. This AGN fraction is correlated with host galaxy mass, and we find an excess of AGNs in mergers (relative to non-mergers) for major mergers (excess by a factor of 1.8) and minor mergers (excess by a factor of 1.5) in the galaxy stellar mass bin $11 < \log (M_* / M_\odot) < 11.5$ (Figure~\ref{fig:fagn}).  Overall, the AGN excess trends somewhat stronger in major mergers than in minor mergers.  

3. When we sort by merger stage, we find the highest AGN excess (relative to non-mergers) for post-coalescence mergers in the highest stellar mass galaxies in MaNGA ($11.5 < \log (M_* / M_\odot) < 12$).  This factor of 2.4 excess has large error bars, and is also consistent  with no excess (Figure~\ref{fig:fagn_mass_stage}).

4. We find no evidence for different merger types or merger stages preferentially triggering higher luminosity AGNs.  AGNs in major mergers and minor mergers have comparable luminosities to AGNs in non-mergers (Figure~\ref{fig:lbol_mass}), and AGNs in early-stage, late-stage, and post-coalescence mergers also have similar luminosities to AGNs in non-mergers (Figure~\ref{fig:lbol_mass_stage}).  

5.  There is no correlation of the merger fraction, which is the ratio of  the number AGNs in mergers to the number of AGNs, with Eddington ratio (Figure~\ref{fig:f_lbolmass}).  We find that an AGN is $\sim$equally like to be found in a merger, regardless of the accretion rate, indicating that galaxy mergers do not enhance the efficiency of accretion onto SMBHs.

In conclusion, we find that AGNs with bolometric luminosities $\log(L_{bol}/ \mathrm{erg} \, \mathrm{s}^{-1}) > 44.4$ are more likely to be triggered or enhanced in galaxy mergers than in non-mergers, for galaxies of stellar mass $\sim 10^{11-12}  \, M_\odot$ in MaNGA.  Although AGNs are more likely to be present in these galaxy mergers, the luminosities and accretion rates of these AGNs are, on average, no different than those of AGNs in non-merging galaxies.  Observations of the molecular gas content of the host galaxies could further clarify whether there is correspondingly no difference in the gas available for accretion onto SMBHs in the merging galaxies as opposed to the non-merging galaxies.

\acknowledgements J.M.C. is supported by NSF AST-1714503 and NSF AST-1847938. R.N. is supported by NSF AST-1714503.  J.N. is supported by NSF AST-1847938.  A.M.S. is supported by a National Science Foundation MPS-Ascend Postdoctoral Research Fellowship under Grant No. 2213288.

Funding for the Sloan Digital Sky Survey IV has been provided by the Alfred P. Sloan Foundation, the U.S. Department of Energy Office of Science, and the Participating Institutions. SDSS-IV acknowledges support and resources from the Center for High-Performance Computing at the University of Utah. The SDSS web site is www.sdss.org.

SDSS-IV is managed by the Astrophysical Research Consortium for the  Participating Institutions of the SDSS Collaboration including the Brazilian Participation Group, the Carnegie Institution for Science, Carnegie Mellon University, the Chilean Participation Group, the French Participation Group, Harvard-Smithsonian Center for Astrophysics, Instituto de Astrof\'isica de Canarias, The Johns Hopkins University, Kavli Institute for the Physics and Mathematics of the Universe (IPMU) / University of Tokyo, the Korean Participation Group, Lawrence Berkeley National Laboratory, Leibniz Institut f\"ur Astrophysik Potsdam (AIP),  Max-Planck-Institut f\"ur Astronomie (MPIA Heidelberg), Max-Planck-Institut f\"ur Astrophysik (MPA Garching), Max-Planck-Institut f\"ur Extraterrestrische Physik (MPE), National Astronomical Observatories of China, New Mexico State University, New York University, University of Notre Dame, Observat\'ario Nacional / MCTI, The Ohio State University, Pennsylvania State University, Shanghai Astronomical Observatory, United Kingdom Participation Group, Universidad Nacional Aut\'onoma de M\'exico, University of Arizona, University of Colorado Boulder, University of Oxford, University of Portsmouth, University of Utah, University of Virginia, University of Washington, University of Wisconsin, Vanderbilt University, and Yale University.  This project makes use of the MaNGA-Pipe3D dataproducts. We thank the IA-UNAM MaNGA team for creating this catalogue, and the Conacyt Project CB-285080 for supporting them.

\bibliographystyle{apj}

\end{document}